# A Pedagogical Study Of The Oyibo's Grand Unification Theorem With Realization Of Some Standard Equations

*Godfrey E. Akpojotor and Myron W. Echenim*
Department of Physics, Delta State University, Abraka 331001, Nigeria
E-mail: akpogea@deltastate.edu.ng; Phone: +234806 660 2677

**Abstract**
The God Almighty Grand Unification Theory (GAGUT) proposed by Oyibo to unify all known forces in nature and other possibly unknown force fields has remained controversial not just because of its ambitious claims but also because of its unconventional mathematical approach. He has adopted the mathematical approach from his experience at solving the Navier Stokes equations in fluid mechanics using invariance of an arbitrary function under a group of conformal transformations. However, this esoteric approach resulted in a sound mathematical formulation for the modelling philosophy of GAGUT which is that since the fundamental characteristic of the universe is motion and motion can only be provided by force, then the universe could be viewed as a large force field. Oyibo then represented the conservation of this large force field at a given space time point in the universe by a set of generic equations from which he obtained his generic solutions whose specific applications depend on the initial/boundary conditions and other physical constraint conditions. An important achievement of the Oyibo's methodology is that modelling with it is reduced to algebraic operations rather than differential equations for the most parts in previous methodologies. With this understanding from pedagogically studying the modelling philosophy and mathematics of the GAGUT, we have been able to recover from it simple standard equations such as in the Fermat principle for geometric optics. This is encouraging and therefore supports the possibility to recover more results and also to provide new ones, thereby supporting the GAGUT as a potential candidate for a grand unification theory.

## INTRODUCTION

Nigeria is celebrating her golden jubilee of independence this year and it is therefore necessary to revisit the work of the Nigerian born Professor of Mathematics of OFFPPIT Institute of Technology, New York, Dr Gabriel Audu Oyibo, who is one of her most celebrated scientists in recent times. Professor Oyibo was born in Nigeria in 1950, studied Mechanical Engineering at the Ahmadu Bello University in Zaria, Nigeria, before proceeding to the Rensselaer Polytechnic Institute where he got a PhD in Aeronautics and Mathematics in 1981. He came into lime light when in 1999 the Time Press New York announced that he had written an article entitled Generalized Mathematical Proof of Einstein's theory using a New Group Theory [1] which he later advanced into the God Almighty Grand Unification Theory (GAGUT) in two books [2]. Though only few scientists in Nigeria were attracted to it then, his nomination for the Nobel Prize and his visit to Nigeria in November, 2004, exposed him but not his work to Nigeria. For the visit was well publicized to the admiration of many of us in the scientific community and therefore gladly hoped the needed revolution in science and technology (S & T) was about to commence especially as the media went agog elevating him to a legendary status while from Aso Rock to his home town, he was warmly received. This include his recognition by the Nigerian Federal Government as a Mathematical Genius which was inscribed on a Nigerian Postage Stamp that was issued in 2005 and his recognition by the Nigerian Senate through the Senate Motion No. 151 page 320 presented in the Federal Republic of Nigeria Order Paper on Tuesday, 15th March, 2005. He was also Knighted by the Attah of the Igala Kingdom as the Amokidojo (translates into English as "Genius from Within") in 2004 during the visit after his state Governor then had earlier led a delegation to visit him in New York.

One intended benefits of the Oyibo lecture tour of his GAGUT packaged by the National University Commission (NUC) was to popularized and probably attract researchers in the country to the GAGUT. In other words, his visit should have galvanized interest in his work and probably activate a group or groups in the country to start research in his work. Furthermore, there should have been topical seminars or conferences to discuss the work. All these would have helped to instigate a scientific revolution in the country and win more young people into the science. About six years now, this has not happened. To the best of our knowledge, outside Animalu who reviewed the work in Refs. [3,4] and a contributed chapter in Ref. [5], it is only my group that has made presentations at conferences and also contributed to the GAGUT literature in Nigeria [6]. One of the reasons [7] for the lack of more studies of GAGUT in the country may be due to the unconventional mathematical methodology introduced by Oyibo. This methodology has been adopted from his experience at solving the Navier Stokes equations in fluid mechanics using invariance of an arbitrary function under a group of conformal transformations [1]. As pointed out by Animalu, the first problem in understanding GAGUT emanates from the lack of direct relationship between this conformal transformations and the usual characterization of conformal invariance or symmetry in analytical projective space-time geometry as well as relativistic quantum field theory [4]. Therefore Animalu provided this missing link by demonstrating how to realize other definitions of the conformal group of transformations within the purview of GAGUT. He was then led to the conclusion in both reviews [3,4] that just as the Minkowskian geometry is the important approach to understanding the Einstein's special relativity theory, projective geometry [8] is the key approach to understanding GAGUT.

Interestingly, Oyibo envisaged the problem of subjective interest on his unconventional methodology and appealed that [1,2]:

*Human experience seems to have demonstrated that under difficult circumstances such as the ones that surround the search for the Unified Force Field Theory, it is critical for one to be open-minded in one's investigation and analysis or even expectations. This reminder to readers is provided to partially prepare them for the coming presentation of the new methodology described in his book. The new methodology would seem to be drastically or significantly different from … the methodology that*

*readers are familiar with or even expect to consider to be the kind of methodologies that belong in this realm of research.*

It is the philosopher, John Dewey, who once asserted that [9], *"Every advance in science has issued from a new audacity of imagination."* This has been exemplified by a number of major scientific advancements by the introduction of esoteric approaches or concepts to solve some difficult problems in most fields of studies. For example, it is a textbook knowledge that the problem which led to the birth of quantum physics was the introduction of the revolutionary idea of quantization by Max Planck to formulate the blackbody radiation law [10]. Another example was in the early development of relativistic quantum mechanics for the electron, wherein the Klein-Gordon theory was considered the best that could be achieved by most contemporary researchers in this field even though there were discrepancies between it and the general principle of quantum mechanics such as its non-positive definite probability density and the presence of symmetry between negative and positive energies. By introducing two valued quantities now known as spinors to get away from tensors which he believed were inadequate then to develop a relativistic quantum theory, Dirac obtained his celebrated theory of relativistic electron [11]. According him [12],

*Those people who were too familiar with tensors were not fitted to get away from them and think up something more general, and I was able to do so only because I was more attached to the general principle of quantum mechanics than to tensors....One should always guard against getting too attached to one particular line of thought*

In our opinion, therefore, it will be very necessary to consider the Oyibo's work with open-mindedness within the general philosophy of a grand unified theorem rather than its deviation from conventional methodologies. This is the purpose of this current pedagogical study of GAGUT and it is planned as follows. In sec. II, we will review the modelling philosophy and statement of the problem of GAGUT. The interesting observation is that the resulting invariant solutions of the generic equations from GAGUT can explored algebraically. This will be exemplified by showing how the hierarchy of the invariant solution for the case where n = 0 to be developed in sec. III has been used to recover some known simple standard equations of physics [6] in sec. IV. Then we will conclude in sec. V.

II. REVIEW OF THE PHILOSOPHY AND STATEMENT OF THE PROBLEM OF GAGUT

The grand unification theories are proposed to unify all known forces in nature such as the four major force fields namely gravitation, electromagnetism, strong and weak forces [13-15] and other possibly unknown force fields [1,2]. In other words, these theories can account for almost every known form of matter and force and possibly the ones that are not yet known, conceivable and non-conceivable. It is believed by some physicists that the successful achievement of such a Theory of Everything (TOE) will lead to the end of physics or at least the beginning of the end [14]. Historically, the Greek were the first to propose that all phenomena of nature can be explained through four 'elements': fire, earth, air and water. The Greek later postulated the idea of the atom as the smallest indivisible chunk of matter. This atom has been observed to constitute a structure as it is made up of electron, proton and neutron. Though both the proton and neutron in turn have their own internal structures as they constitute smaller particles, the four known fundamental forces in nature namely electromagnetic, strong, weak and gravitational forces can be analyzed using the constituents of the atom: the electromagnetic force is that between the electron and proton, the strong force is between proton and neutron, the weak force is between pairs of protons and neutrons and the gravitational force is that between any of these chunks of matter [16,17].

The motivation to unite the four forces emanates from the unification of the originally separate forces of electricity and magnetism as an electromagnetic force. Scientifically, Einstein began the quest for a unified force field theory when he attempted to incorporate electromagnetism into his General Relativity Theory. As it is now well known in Textbooks, Einstein mathematical framework for his Special Theory of Relativity is the Lorentz group of linear coordinate transformations and by generalizing these transformations to include non-linear cases, he was able to set up the mathematical framework for the General Relativity Theory. It is therefore this methodology of general coordinate transformations that Einstein attempted to unify electromagnetic and gravitational force fields. Therefore, most other workers in the search for a unified force field have adopted the Einstein methodology or modifications of it and they have not been successful [15]. One of the few works that have attempted the unification problem from a different philosophy is the approach by Salam, Weinberg and Sheldon to unify the electromagnetic and weak forces which won the 1979 Nobel Prize in Physics [16-18]. By assuming that the inclusion of gravitational force field is the hardest problem in the quest of unification of forces, these workers deliberately avoid gravity and concentrated on unifying the electromagnetic force with the nuclear forces of weak and strong interactions. However, unlike the electroweak force which has been confirmed experimentally, the experiments to confirm their theory of unification of the electromagnetic and strong forces are still controversial [16, 18-20]

The Oyibo methodology is esoteric as already stated and this is based on his perception of some previous works in the quest for GUT and what he now conceives the GUT to mean [1]:

*A physically sound or credible set of mathematical equations from which to determine or formulate the Grand Unified Force Field Theory comprising of the four known forces in the universe which are the gravitational, the electromagnetic and the nuclear forces of strong forces and weak forces as well as other forces which may not have already been discovered.*

To obtain such set of equations, Oyibo model the problem as follows:

*'The most fundamental characteristic of the universe is motion. This fundamental thing about the universe being motion can be basically derived from the fact that, the material universe is made up of atoms consisting of electrons rotating around the atomic nucleus perpetually, plus planets motions and solar systems motion and the motion of galaxies, etc. This gives us the understanding that the universe is basically characterized by motions. Therefore since motion can only be provided by force, the universe could be viewed as a large force field.'*

Oyibo then represented the conservation of this large force field at a given space time point in the universe by demanding that an arbitrary function G given by

$$G = G(Y^1, Y^2 ... Y^p) \qquad (2.1)$$

should be conformally invariant under the group of transformation:

$$T_k : Y^i = f^i(y^1...y^p, k) \qquad (2.2)$$

if $T_k$ is the group of transformations and

$$G = G(Y^1, Y^2 ... Y^p) = F(y^1...y^p, k).(y^1, y^2...y^p) \qquad (2.3)$$

where $F(y^1...y^p, k)$ is a function of $y^i$ and $k$ the single group parameter.

Now this group of transformations are to obey a new set of group laws and possess a new form of group parameters [21,22]. The argument of Oyibo is that in the final analysis, what establishes the integrity of this methodology is not so much the group laws or group defined parameters but the end results or final conclusion reached. With this conjection, Oyibo derived a set of conservative equations

$$(G_{0n})_t + (G_{1n})_x + (G_{2n})_y + (G_{3n})_z = 0, \qquad (2.4)$$

where n = 0, 1, 2, 3, 4.

Eq. (2.4) can be expressed in the Einstein-like form of conservative equations:

$$G_{mn} = 0. \tag{2.5}$$

This is the Oyibo generic (meaning the specific nature is determined by the initial/boundary conditions and other physical constraint conditions) conservation equation which is an arbitrary function of space and time coordinates (x,y,z,t), velocities $(\dot{x}, \dot{y}, \dot{z})$, density $(\rho)$, fluid or gas viscosity $(\mu)$, temperature (T), pressure (P), etc:

$$G_{mn} = G_{mn}(x, y, z, t, \dot{x}, \dot{y}, \dot{z}, \rho, \mu, T, P, ....). \tag{2.6}$$

When the transformations in Eq. (2.2) is generalized to a system of partial differential equations of order n given by

$$G_j\left[x^1, x^2, ...x^p, y^1, y^2, ...y^q, ...., \frac{\partial^n y^1}{(\partial x^1)^n} ... \frac{\partial^n y^q}{(\partial x^p)^n}\right] = 0 \tag{2.7}$$

and is conformally invariant under the transformations $T_k^n$, then the generic solutions to Eq. (2.4) is

$$\eta_n = g_{n0} t^{n+1} + g_{n1} x^{n+1} + g_{n2} y^{n+1} + g_{n3} z^{n+1} \tag{2.8}$$

where $\eta_n$ is the absolute invariant of the subgroup of transformations for the independent coordinate variables and $g_{n0}, g_{n1}, g_{n2}$ and $g_{n3}$ are metric parameters.

The Oyibo`s generic equation in Eq. (2.5) can be recast into matrix form for $m, n = 0,1,2,3$ say,

$$\begin{bmatrix} G_{00} & G_{01} & G_{02} & G_{03} \\ G_{10} & G_{11} & G_{12} & G_{13} \\ G_{20} & G_{21} & G_{22} & G_{23} \\ G_{30} & G_{31} & G_{32} & G_{33} \end{bmatrix} \tag{2.9}$$

The subgroup of transformation for the coordinate's variables is characterized by the relationship $T_n : X_n : Y_n : Z_n = ct : x : y : z$. Therefore the hierarchy of the Oyibo`s invariant solution for Eq.(2.9) has the following forms

$$\eta_0 = g_{00} ct + g_{10} x + g_{20} y + g_{30} z \tag{2.10a}$$

$$\eta_1 = g_{01}(ct)^2 + g_{11} x^2 + g_{21} y^2 + g_{31} z^2 \tag{2.10b}$$

$$\eta_2 = g_{02}(ct)^3 + g_{12} x^3 + g_{22} y^3 + g_{32} z^3 \tag{2.10c}$$

$$\eta_3 = g_{03}(ct)^4 + g_{13} x^4 + g_{23} y^4 + g_{33} z^4 \tag{2.10d}$$

In his review [3], Animalu demonstrated how to construct the realization of the hierarchy of solutions of the generic equations for n = 0, 1, 2, 3, 4. Our goal here is to explore the hierarchy of the invariant solution for the case where n = 0 and try to recover some known standard equations of physics.

## III. OBTAINING THE SPACE-TIME INVARIANCE FOR THE GENERIC SOLUTION FOR n = 0

McConnell states [23]

*Let A be a point whose coordinates are $x^r$ and let R be any neighboring point with coordinate $x^r + dx^r$. If we donate the infinitesimal distance AR by ds, which is also called the element of the path, a 4-dimensional form for a physical metric is stated as*

$$(\partial s)^2 = \partial s^2$$
$$= h_0 g_0 (\partial x^0)^2 + h_1 g_1 (\partial x^1)^2 + h_2 g_2 (\partial x^2)^2 + h_3 g_3 (\partial x^3)^2 \quad (3.1)$$

where the $x^0, x^1, x^2, x^3$ denotes distinct variable that are used to denote a point in space-time [1]

The space-time of a physical event can be described as a real and smooth manifold $M_D$ with coordinates $x_i = 0, 1, 2, 3$ while $(\partial s)^2$ is the infinitesimal interval between two infinitesimal points on $M_D$ and which eventually corresponds to the temporal and spatial world-line in the external world

$$(\partial s)^2 = (\partial t)^2 + (\partial r)^2 \quad (3.2)$$

where $(\partial r)^2 = (\partial x)^2 + (\partial y)^2 + (\partial z)^2$ represent the space coordinates of world-line in the manifold $M_D$ and $(\partial t)^2 = (c \partial t)^2$ represent the time part of the world-line of the manifold $M_D$

The challenge at hand is to be able to show that when n = 0, the invariant solution ($\eta_0$) given by Eq. (2.10a) and $(\partial s)^2$ are equivalent. In search of the transformation law, we will be looking at the equation of a plane through the point A with position vector a and perpendicular to a unit position vector $\hat{n}$ (see Fig. 1):

$$(r - a) \cdot \hat{n} = 0. \quad \text{3.3a)}$$

This follows since the vector joining A to the general point R with the position vector r is r-a and r will lie in the plane, if the vector is perpendicular to the normal to the plane

$$r \cdot \hat{n} = a \cdot \hat{n}. \quad (3.3b)$$

Eq. (3.3a) can be recast into the form of $r \cdot \hat{n} = d$

$$lx + my + nz = d \quad (3.3c)$$

where the unit normal to the plane is $\hat{n} = li + mj + nk$ and $d = a \cdot \hat{n}$ is the perpendicular distance of the plane from the origin.

The equation of a plane containing points a, b, c is

$$r = a + \lambda(b - a) + \mu(c - a) \quad (3.3d)$$

A more symmetric form of the equation will be of the form [24]

$$r = \alpha a + \beta b + \gamma c \quad (3.3e)$$

where $+\beta + \gamma = 1$.

Now let's consider a curve $r(s)$, parameterized by an Arc length s from some point on the curve, if we write the length of the elemental path $(\partial s)^2$ in the form of equation

$$(\partial s)^2 = \alpha(\partial x)^2 + \beta(\partial y)^2 + \gamma(\partial z)^2, \quad (3.4a)$$

by including the time component into Eq.(3.4a), the resulting equation becomes

$$(\partial s)^2 = \varepsilon(\partial t)^2 + \alpha(\partial x)^2 + \beta(\partial y)^2 + \gamma(\partial z)^2 \quad (3.4b)$$

where $\varepsilon, \alpha, \beta, \gamma$ are constraining constants; if $\varepsilon = \alpha = \beta = \gamma = 1$, Eq. (3.4b) becomes

$$(\partial s)^2 = (c\partial t)^2 + (\partial x)^2 + (\partial y)^2 + (\partial z)^2 \quad (3.4c)$$

which can also be expressed in the familiar form of the equation of metric as

$$(\partial s)^2 = (\partial x^0)^2 + (\partial x^1)^2 + (\partial x^2)^2 + (\partial x^3)^2 \quad (3.4d)$$

where $x^0 : x^1 : x^2 : x^3 = ct : x : y : z$.

The expression for a general infinitesimal vector displacement $\partial r$ is given by [25]

$$\partial r = \frac{\partial r}{\partial x}\partial x + \frac{\partial r}{\partial y}\partial y + \frac{\partial r}{\partial z}\partial z$$

$$= i\partial x + j\partial y + k\partial z \quad (3.5a)$$

where $\frac{\partial r}{\partial x} = i, \frac{\partial r}{\partial y} = j$ and $\frac{\partial r}{\partial z} = k$.

We note that a scalar product operation does not alter the geometric character of the function to which it is applied, the scalar product of $\partial r$ with $\nabla \eta_0$ will give us

$$\nabla \eta_0 \cdot \partial r = \left(i\frac{\partial \eta_0}{\partial x} + j\frac{\partial \eta_0}{\partial y} + k\frac{\partial \eta_0}{\partial z}\right) \cdot (i\partial x + j\partial y + k\partial z)$$

$$= \frac{\partial \eta_0}{\partial x}\partial x + \frac{\partial \eta_0}{\partial y}\partial y + \frac{\partial 0}{\partial z}\partial z = \partial \eta_0 \quad (3.5b)$$

This is the infinitesimal change in $\eta_0$ going from $r \to r + \partial r$, since r depends on x, y, z such that $r(x, y, z)$ defines a space curve, that is, the total derivative of $\eta_0$ with respect to x, y, z along the curve is given by

$$\nabla \eta_0 \cdot \partial r = \partial \eta_0 . \quad (3.5c)$$

A careful inspection of Eq. (3.5c) shows that it is the differential form of the spatial coordinates of equation (2.10a). We now will rewrite Eq. (3.5c) in the form which will now include both the time and space component.

$$\partial \eta_0 = \frac{\partial \eta_0}{\partial t}.(c\partial t) + \nabla \eta_0 . \partial r$$

$$\partial \eta_0 = \frac{\partial \eta_0}{\partial t} c\partial t + \frac{\partial \eta_0}{\partial x}\partial x + \frac{\partial \eta_0}{\partial y}\partial y + \frac{\partial 0}{\partial z}\partial z \qquad (3.5d)$$

From the earlier definition of $\varepsilon, \alpha, \beta, \gamma$,

$$\frac{\partial \eta_0}{\partial t} = g_{00} = \varepsilon, \quad \frac{\partial \eta_0}{\partial x} = g_{10} = \alpha, \quad \frac{\partial \eta_0}{\partial y} = g_{20} = \beta, \quad \frac{\partial \eta_0}{\partial z} = g_{30} = \gamma$$

$$\partial \eta_0 = c\partial t + \partial x + \partial y + \partial z. \qquad (3.6)$$

By applying the conditions for Orthogonality [25]

$$(\partial \eta_0)^2 = (c\partial t)^2 + (\partial x)^2 + (\partial y)^2 + (\partial z)^2. \qquad (3.7)$$

It is easy to observe by comparing Eq. (3.4c) and (3.7) that the differential form of the invariant solution $(\partial \eta_0)^2$ is equivalent to $(\partial s)^2$

$$(\partial s)^2 = (\partial \eta_0)^2. \qquad (3.8)$$

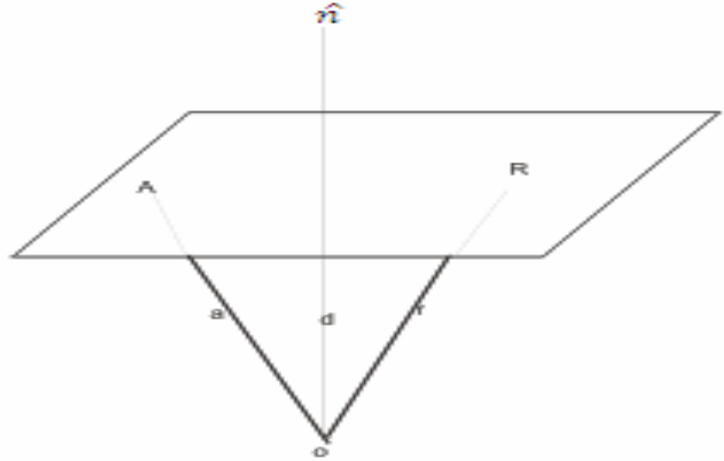

**Fig. 1:** A plane through the point A with position vector a and perpendicular to a unit position vector $\hat{n}$.

## IV    RECOVERY OF RESULTS IN GEOMETRIC OPTICS

Geometric Optics gives only an approximation for small wavelength of D`Alembert equation as Arnold Summerfield and Iris Runge demonstrated in 1911 but yield the exact propagation law of the electromagnetic waves front, quite independently of the structure wave, a fundamental result which follows from the theory of the characteristic manifold of the partial differential equation of the second order. The law dictating electromagnetic radiation propagation in a vacuum is the basic tenet of special relativity and the elemental path $(\partial s)^2$ of the travelling electromagnetic wave includes the results of geometric optics.

From Eq. (3.4c) and (3.7), we have showed that the metric of the square between the neighbouring points in space-time which is invariant is equivalent to the differential form of the Oyibo`s invariant solution when n = 0 (Eq. (2.10a)), that is,

$$(\partial s)^2 = (\partial \eta_0)^2$$
$$= g_{00}(c\partial t)^2 + g_{10}(\partial x)^2 + g_{20}(\partial y)^2 + g_{30}(\partial z)^2 \qquad (4.1)$$

Therefore, if we interpret Eq. (4.1) as an expression of space-time interval in Minkowski manifold that would refers to a system of general coordinates, then we can recover the results of geometric optics in a vacuum as it is described by special relativity. This is the application we now turn to.

**Fermat Principle in optics**

The Fermat principle in optics states that [24]

*A ray of electromagnetic wave travelling through a medium of variable refractive index will follows a path such that the total optical path length, P is stationary,* that is,

$P$ = physical path x refractive index

Now if assume that the plane in Fig. 1 has an interface that is constant at y which separates it into two regions with refractive indices $n_1$ and $n_2$, then the element of physical path length of the electromagnetic wave, a ray of light say, is

$$\partial s = (1 + y'^2)^{1/2} \partial x. \qquad (4.2)$$

Hence its total optical path length as it passes through the pints A and R is

$$P = \int_A^R n(y)(1 + y'^2)^{1/2} \partial x. \qquad (4.3)$$

By setting $g_{00}$ and $g_{30}$ to be zero and $g_{10}$ and $g_{20}$ to be unity in Eq. (4.1), then Eq.(4.3) can be also be expressed using the differential form of the Oyibo's invariant solution for n = 0 as

$$P = \int_A^R n(y) \partial \eta_0. \qquad (4.4)$$

Applying the appropriate method of calculus of variation [24] to resolve Eq. (4.3), the general solution of Eq. (4.4) will be

$$n(y) \sin \theta = constant \qquad (4.5)$$

with $n(y)$ and $\theta$ varying individually.

**Case I**

When $n(y)$ and $\theta$ varies, then Eq. (4.6) can be expressed as

$$n(y)_1 \sin \theta_1 = n(y)_2 \sin \theta_2 \qquad (4.6)$$

which is the Snell `s Law of refraction at an interface.

**Case II**

When $n(y)$ is unity, that is,

$$\frac{n(y)_1}{n(y)_2} = 1, \tag{4.7}$$

we will recover the special case of non-bending light path at an interface having the same refractive indices in both regions,

$$\frac{\sin\theta_1}{\sin\theta_2} = 1. \tag{4.8}$$

## V. SUMMARY AND CONCLUSION

We have pedagogically studied the modelling philosophy and mathematics of the Oyibo GAGUT and our understanding is that it is a theorem, as such it is simply a rule or principle that can be proved mathematically and this is what Oyibo has done. This means there is no possibility of errors logically or geometrically in GAGUT. Therefore, any experiment which fails to verify the GAGUT equation is deemed to have been done incorrectly since GAGUT is a Theorem. Thus one can understand why it has not been faulted after it has been reviewed extensively by some of the greatest mathematicians and also by the American Mathematical Society. Now it a common knowledge that mathematics is the language with which the physicist communicate his ideas compactly, economically and beautifully. However, it is not every mathematics that is applicable in physics. This is why in every physics curriculum, there are courses on mathematical methods in physics [24,25] where future physicists must learn those mathematics that have applications in physics as well as learn their limitations. Following this line of thinking, the application of a theorem in physics requires some good knowledge of that particular area of physics. Thus the realization or recovery of previous results as well as predicting new ones requires an unbiased study of GAGUT and the relevant physics knowledge. To buttress this point, we have been able to recover both the Snell's law and the law of refraction from the generic solution for the motion of wave, $\eta_0$. This is encouraging and therefore supports the possibility that with more works, it may be possible to recover previous results from GAGUT and also some of the predictions of Oyibo. This conjecture is in line with the assertion Einstein once made that whether one observed a thing or not depends on the theory which one used [26]. In other words, it is the theory that decides what can be observed since observation is the connection constructed from the phenomena and our realization. His reason being that any reasonable theory will besides all the things that can immediately be observed from it, give the possibility of observing other things more indirectly. We think this is the goal of the Oyibo's GAGUT.

One of the controversial predictions of Oyibo from GAGUT is that hydrogen which he also called Africanium is the only element in nature and hence the building block of matter [1,2]. This possible single element theory is in line with the principle of nucleosynthesis which is a process of heavyweight element formation by fusion of two lightweight elements at extremely high temperature and pressure found in stars and supernovae [27]. A common example is the fusion hydrogen atoms to form helium as in the interior of the sun which in turn can combined to produce carbon which become the fuel for producing even heavier elements such as oxygen [28]. The implication of the single element theory is that nucleosynthesis which is currently man's major means of element formation in line with the Big Bang theory, will be reduced to understanding how the Hydrogen atom can combine to form all the other 117 previously called elements in the periodic table as its nuclear compounds and to use this idea to predict new ones. This is why Oyibo predicts that this approach will make the study of science 118 times simpler [1].

Another controversial prediction of the GAGUT has to do with the claim by Oyibo that it can be used to solve various man's problems including those in health and economics. While this may seems overambitious, one must not lose sight of the impact of the Einstein mass-energy equation, $E = mc^2$. This world most famous equation is believed to have revolutionalized physics, redefine strategic arms, and promises to transform our economy and environment with plentiful, clean energy [29]. It is therefore not naïve to postulate that if the extension of this equation which is one of the salient conclusions from GAGUT that mass can be transformed not only into energy but also into momentum is verified, then GAGUT may also hold the possibility of extending the promises of the Einstein mass-energy equation.

Before concluding, it is pertinent to remind readers that an inapplicable theory, even a wrong theory, can be interesting, inspiring and may turn out to be useful in another setting [30]. Further, the rejection of a theory by some established workers in the field does not translate into its total failure. For example [18], when Salam proposed his left-right symmetry violation in weak nuclear forces to account for the zero mass of the neutrino to Peierls, the later rejected it saying, "*I do not believe left-right symmetry violation in weak nuclear forces at all. I would not touch such ideas with tongs.*" Discouraged but not disillusioned, he took the paper to Pauli who is acclaimed to be the father of neutrino and he met rejection there too. First, Villars of MIT who was visiting Pauli the same day he took the paper to him, returned his verdict the following day, "*Give my regards to my friend Salam and tell him to think of something better.*" Similarly, Pauli rejected the work after an elaborate review. However, he made some useful comments on the entire work which Salam later considered and with persistence, the left-right symmetry spontaneous breaking became the paradigm for the realization of the electroweak force.

GAGUT is still in the controversial stage and therefore needs more studies to recover more previous results and to predict new ones. This is the global challenge especially to the Nigerian physics community and to all those including the media and government and their agencies that went agog when Oyibo visited the country with his GAGUT. For it is a common knowledge today that the accomplishments of the scientists of a country do play a major role on the image of the country.

**Acknowledgement**

We appreciate the very inspirational discussion with Professor A.O.E. Animalu and also for making available to us his papers. We also acknowledge the useful discussion with Professor Amagh Nduka. This work is supported in part by ICBR and also by AFAHOSITECH.